\documentclass[aps,preprint]{revtex4}
\usepackage{amsmath}
\usepackage{epsfig}

\begin{document}

\title{THE LORENTZ INTEGRAL TRANSFORM (LIT) METHOD}

\author{Winfried Leidemann}

\affiliation{Dipartimento di Fisica, Universit\`a di Trento\\
and Istituto Nazionale di Fisica Nucleare, Gruppo Collegato di Trento\\
Via Sommarive 14, I-38100 Trento, Italy\\
leideman@science.unitn.it}

\date{\today}

\begin{abstract}
The LIT approach is reviewed both for inclusive and exclusive reactions.
It is shown that the method reduces a continuum state problem to a bound-state-like
problem, which then can be solved with typical bound-state techniques. The LIT
approach opens up the possibility to perform {\it ab initio} calculations of reactions also
for those particle systems which presently are out of reach in conventional approaches
with explicit calculations of many-body continuum wave functions. Various LIT applications are
discussed ranging from particle systems with two nucleons up to particle systems with seven
nucleons.
\end{abstract}

\maketitle

\section{Introduction}
{\it Ab initio} calculations are a central element of few-body physics. The only input
in such calculations is a well defined Hamiltonian. In nonrelativistic nuclear physics
one then solves the Schr\"odinger -- or equivalent -- equations without introducing
any approximation. In order to calculate cross sections for reactions, where final or
initial states are scattering states in the continuum, one is faced with the problem
of an {\it ab initio} calculation of such continuum states. It is well known that already the 
calculation of a three-body continuum wave function is quite difficult and that today a complete
four-body calculation, with all possible break-up channels open, is out of reach. However,
the problem of a many-body continuum state calculation can be circumvented if one uses
the Lorentz integral transform (LIT) method \cite{ELO94}.
In fact the LIT approach allows the {\it ab initio} calculation of reaction cross sections,
where a many-body continuum is involved, without requiring the knowledge of the generally 
complicated many-body continuum wave function. The scattering problem is reduced to a 
calculation of a localized function with an asymptotic boundary condition similar to a
bound-state wave function. Such an approach was already proposed by Efros in 1985, 
but with the Stieltjes instead of the Lorentz integral transform \cite{Efr85}. However, it 
has been found that the application of the Stieltjes transform is problematic 
since it leads to serious inversion problems \cite{Stieltjes}. 

The LIT method has been applied to various
electroweak cross sections in the nuclear mass range from A=3 to A=7. Among the applications 
are the first realistic {\it ab initio} calculations of the nuclear three- and four-body total
photoabsorption cross sections \cite{ELOT00,Doron06}, as well as of the inelastic neutral current 
neutrino scattering off $^4$He \cite{GN07}. In addition first {\it ab initio} calculations have 
been  performed for the total photoabsorption cross sections of $^{4,6}$He and $^{6,7}$Li with 
semirealistic forces \cite{4-body,6-body,7-body}. Other applications were carried out for the
inelastic inclusive electron scattering cross section (see {\it e.g.} 
\cite{ee'97,ee'05,ee'08}). Besides inclusive electroweak reactions
also LIT calculations of exclusive reactions have been performed, namely for 
$^4$He$(\gamma,n)^3$He and $^4$He$(\gamma,p)^3$H \cite{Sofia1}, $^4$He$(e,e'p)^3$H \cite{Sofia2},
and $^4$He$(e,e'd)d$ \cite{Diego}. Further applications and a detailed 
description of the LIT method are presented in a recent review article \cite{ELOB07}.

The paper is organized as follows. In section II the general form of inclusive and exclusive 
cross sections of a particle system induced by an external probe is outlined. Sections 
III and IV describe the LIT approach for inclusive and exclusive response functions, respectively.
In section V various LIT applications are discussed.

\section{Structure of electroweak cross sections}

Perturbation-induced reactions can be divided in inclusive and exclusive processes. In the 
former case the final state of the particle system is not observed. In the latter case
the final state is at least partially observed, like {\it e.g.} in an $(e,e'p)$ reaction,
where besides the scattered electron also an outgoing proton with energy $E_p$
and scattering angle $\Omega_p=(\theta_p,\Phi_p)$ is detected.

Inclusive cross sections of perturbation-induced reactions have the following general
form (see {\it e.g.} \cite{textbook})
\begin{equation}\label{incl_cross}
{\frac {d^2\sigma}{d\omega d\Omega_{ext}}} = \alpha_{\rm ext} 
\sum_{i=1}^M f_i(\omega,q,\theta_{ext}) R_i(\omega,q) \,,
\end{equation}
where $\omega$ and $q$ and are energy and momentum transfer of the external probe
to the particle system, $\Omega_{ext}=(\theta_{ext},\phi_{ext})$ denotes the scattering 
angle of the external probe, $\alpha_{\rm ext}$ is a constant characteristic for the
external probe, and $f_i$ are kinematic functions.
The functions $R_i$ describe the various responses of the particle system to the external
probe and thus contain information about the dynamics of the particle system.
They are defined as follows
\begin{equation}\label{response}
R_i(\omega,q)=\sum\!\!\!\!~\!\!\!\!\!\!\!\!\int _f\,\,
|\langle f|\Theta_i| 0 \rangle |^2 \delta(\omega -(E_f-E_0)) \, .
\end{equation}
Here $E_0$ and $|0\rangle$ are ground state energy and wave function of the
particle system under consideration, $E_{f}$ and $|f\rangle$ denote final state energy 
and wave function of the final particle system, and $\Theta_i$ 
is the operator inducing the response function $R_i$. 

As an example for an exclusive cross section we consider the $(e,e'p)$ case, here one has
\begin{equation}\label{exl_cross}
{\frac {d^3\sigma}{d\omega d\Omega_{e} d\Omega_{p}}} = \alpha_{\rm ext}
\sum_{i=1}^N f_i(\omega,q,\theta_{e}) g_i(\phi_p) r_i(\omega,q,\theta_p,E_p) \,,
\end{equation}
where the $\phi_p$ dependence of the cross sections is described by the known
functions $g_i(\phi_p)$.
Exclusive response functions $r_i$ do not have such a simple form as the inclusive
functions $R_i$. For their definition we refer to \cite{textbook}, here we only mention
that transition matrix elements from the ground state $| 0 \alpha \rangle$ to a
specific final state $| f \beta \rangle$, 
\begin{equation}\label{T_i}
T^{\alpha\beta}_{0f,i} = \langle f\beta| \Theta_i | 0 \alpha \rangle  \,,
\end{equation}
are their essential ingredients, where $\alpha$ and $\beta$ stand for additional quantum 
numbers of initial and final state wave functions.

The following relation between the inclusive $R_i$ of (\ref{incl_cross}) and 
the exclusive $r_i$ of (\ref{exl_cross}) holds:
\begin{equation}\label{relation}
R_i(\omega,q) = \int d\Omega_p dE_p g_i(\phi_p) r_i(\omega,q,\theta_p,E_p) \,.
\end{equation}
Note that the number of exclusive response functions is generally greater than the number of 
inclusive ones ($N>M$), since the integration over the azimuthal angle of the outgoing 
particle can yield zero, {\it i.e.} the integration over $\phi_p$ in the $(e,e'p)$ example above.

With additional polarization degrees of freedom for beam and/or target and/or 
outgoing particles many more additional inclusive and exclusive response functions can be 
defined (see {\it e.g.} the deuteron case in \cite{pol_deuteron}).

\section{Calculation of inclusive responses with the LIT method}

As already mentioned, with the LIT method one avoids the explicit calculation of
scattering wave functions. Instead, for the calculation of $R_i$ of (\ref{response}) one 
proceeds in the following way. One first calculates the ground state wave function $|0\rangle$
of the particle system in question. Then one solves the equation
\begin{equation} \label{LITeq}
(H-E_0-\sigma_R - i \sigma_I)| \tilde \Psi_i \rangle = \Theta_i |0\rangle \,,
\end{equation}
where $H$ is the Hamiltonian of the particle system and $\sigma_{R/I}$ are parameters,
whose meaning is explained below. Since the eigenvalues of $H$ have to be real the homogeneous 
version of (\ref{LITeq}) has only the trivial solution $\tilde\Psi_i=0$ and thus (\ref{LITeq})
has a unique solution. In addition, due to the asymptotically vanishing ground state wave 
function $|0\rangle$ also the right-hand-side of (\ref{LITeq}) vanishes asymptotically. 
Therefore, and because of the complex energy $E_0-\sigma_R - i \sigma_I$, $\tilde\Psi_i$ has 
a similar asymptotic boundary condition as a bound state. It means that $\tilde\Psi_i$ is a 
so-called localized function, {\it i.e.} square-integrable with a norm 
$\langle\tilde\Psi_i|\tilde\Psi_i\rangle$. This has very important consequences: even
if the aim is a calculation of a reaction cross section in the continuum, one 
is not confronted with a scattering state problem any more, in fact one needs to apply only 
bound-state techniques for the solution of (\ref{LITeq}).

The key point of the LIT method consists in the fact that the Lorentz integral transform
L$_i$ of the response function $R_i$,
\begin{equation}\label{deflit}
{\rm L}_i(\sigma_R,\sigma_I,q) =
 \int R_i(\omega,q)\,  {\mathcal L}(\omega,\sigma_R,\sigma_I)\, d\omega 
\end{equation}
is related to the norm $\langle\tilde\Psi_i|\tilde\Psi_i\rangle$, which can be obtained from 
the solution of (\ref{LITeq}). In fact one has
\begin{equation}\label{deflit2}
{\rm L}_i(\sigma_R,\sigma_I,q) =
  \langle\tilde\Psi_i(\sigma_R,\sigma_I,q)|\tilde\Psi_i(\sigma_R,\sigma_I,q)\rangle
\end{equation}
($q$ dependence of ${\rm L}_i$ and $\tilde\Psi_i$ will be dropped in the following).
In (\ref{deflit}) ${\mathcal L}$ is a Lorentzian centered at $\sigma_R$ with a width 
$\Gamma=2\sigma_I$:
\begin{equation}
{\mathcal L}(\omega,\sigma_R,\sigma_I)=
{\frac {1} {(\omega -\sigma_R)^2 + \sigma_I^2}}\,.
\end{equation}
Now also the meaning of the parameters $\sigma_{R/I}$ becomes evident: $\sigma_I$ represents 
a kind of energy resolution, while with $\sigma_R$ a given energy range can be scanned.

With the above equations the principle idea of the LIT method can be explained:
one solves the LIT equation (\ref{LITeq}) for many values of $\sigma_R$ and a fixed
$\sigma_I$, calculates L$_i(\sigma_R,\sigma_I=$const), and then one inverts the transform
in order to determine $R_i(\omega,q)$. 

Before coming to the inversion we first want to derive the relation (\ref{deflit2}).
Starting from the definition of L$_i$ in (\ref{deflit}) one has
\begin{equation}
{\rm L}_i(\sigma_R,\sigma_I) = 
\int d\omega {\frac {R_i(\omega,q)} {(\omega -\sigma_R)^2 + \sigma_I^2}} =
\int d\omega {\frac {R_i(\omega,q)} 
   {(\omega -\sigma_R + \sigma_I) (\omega -\sigma_R - \sigma_I)}} \,.
\end{equation}
Using (\ref{response}) and carrying out the integration in $d\omega$ one gets
\begin{eqnarray}
{\rm L}_i &=& \nonumber 
\int d\omega {\frac {\sum\!\!\!\!~\!\!\!\!\!\int_{\,\,f}\,
                     \langle 0| \Theta_i^\dagger | f \rangle
		     \langle f| \Theta_i | 0 \rangle}
	{(\omega -\sigma_R + \sigma_I) (\omega -\sigma_R - \sigma_I)}}
	\delta(\omega-(E_f-E_0))   \\
 &=& \sum\!\!\!\!~\!\!\!\!\!\!\!\int_{\,\,\,f}
    \langle 0| \Theta_i^\dagger (E_f-E_0-\sigma_R+ i \sigma_I)^{-1} | f \rangle
    \langle f| (E_f-E_0-\sigma_R - i \sigma_I)^{-1} \Theta_i | 0 \rangle .
\end{eqnarray}
Then one replaces $E_f$ by the Hamilton operator $H$ and uses the closure property of the
eigenstates of the Hamiltonian 
($\sum\!\!\!\!~\!\!\!\!\!\int_{\,\,f}\,\,|f\rangle\langle f| =1$):
\begin{eqnarray}
{\rm L}_i &=& \nonumber
            \langle 0| \Theta_i^\dagger (H-E_0-\sigma_R+ i \sigma_I)^{-1} 
     (H-E_0-\sigma_R- i \sigma_I)^{-1} \Theta_i | 0 \rangle \\ 
 &\equiv& \langle \tilde\Psi_i | \tilde\Psi_i \rangle
\end{eqnarray}
with
\begin{equation}
|\tilde\Psi_i\rangle = (H-E_0-\sigma_R- i \sigma_I)^{-1} \Theta_i | 0 \rangle \,.
\end{equation}
One sees that relation (\ref{deflit2}) is indeed obtained and that $\tilde\Psi_i$ fulfills
(\ref{LITeq}).

The standard LIT inversion method consists in the following ansatz for the response function 
\begin{equation}
R_i(\omega',q) = \sum_{m=1}^{M_{\rm max}} c_m \chi_m(\omega',\alpha_j) \,,
\label{sumr}
\end{equation}
here the argument $\omega$ of $R_i$ is replaced by $\omega'=\omega-\omega_{th}$,
where $\omega_{th}$ is the break-up threshold of the reaction into the continuum.
In case of LIT contributions attributed to bound states due to, {\it e.g.}, elastic
transitions, one can easily subtract such contributions in order to obtain an ``inelastic''
LIT (see \cite{ELOB07}).
The $\chi_m$ are given functions with nonlinear parameters $\alpha_j$.
Normally the following basis set is taken
\begin{equation}
\label{bsetn}
\chi_m(\omega',\alpha_j) = \omega'^{\alpha_1} \exp(- {\frac {\alpha_2 \omega'} {m}}) \,.
\end{equation}
Substituting such an expansion into the right hand side of
(\ref{deflit}) one obtains
\begin{equation}
{\rm L}_i(\sigma_R,\sigma_I) =
\sum_{m=1}^{M_{\rm  max}} c_m \tilde\chi_m(\sigma_R,\sigma_I,\alpha_j) \,,
\label{sumphi}
\end{equation}
where
\begin{equation}
\tilde\chi_m(\sigma_R,\sigma_I,\alpha_j) =
\int_0^\infty d\omega' {\frac {\chi_m(\omega',\alpha_j)} {(\omega'-\sigma_R)^2 + \sigma_I^2}}
\,\,.
\end{equation}
For given values of $\alpha_j$ and $M_{\rm max}$ the linear parameters $c_m$ are determined from
a best fit of L$_i(\sigma_R,\sigma_I)$ of (\ref{sumphi}) to the calculated
L$_i(\sigma_R,\sigma_I)$ of (\ref{deflit2}) for a fixed $\sigma_I$ and a number of
$\sigma_R$ points much larger than $M_{\rm max}$. In addition one should vary
the various nonlinear parameter $\alpha_j$ over a sufficiently large range. The parameter
$\alpha_1$, however, can in general be determined from the known threshold behavior of the
response function. One has to increase $M_{\rm max}$ up to the point that a stable
inversion result is found for some range of $M_{\rm max}$ values, which then can be taken
as final inversion result. Note, however, that a too large value of $M_{\rm max}$ might lead 
to an oscillatory behavior of  $R_i$. The origin for such an unphysical behavior lies in the 
precision of the calculated L$_i$. If the precision is further increased, higher and higher 
$M_{\rm max}$ values can in principle be used in the inversion (see also \cite{Diego2}).
 
One can repeat the whole procedure with a second $\sigma_I$ value.
Of course, the inversion should lead to the same $R_i$ result as with the previous $\sigma_I$. 
The basis set $\chi_m$ can also be modified in order to take into account narrow
structures like resonances (see section V.A). More information concerning the inversion and 
alternative inversion methods are found in \cite{Diego2}.

\section{Calculation of exclusive responses with the LIT method}
   
For the exclusive response function $r_i$ one has to evaluate T-matrix elements of the
type given in (\ref{T_i}). One starts the LIT calculation using the general form of the final
state wave function for the considered break-up channel \cite{Goldb},
\begin{equation}
|\Psi_f^-(E_f)\rangle = |\Phi_f(E_f)\rangle + (E_f-H-i\eta)^{-1} V|\Phi_f(E_f)\rangle \,, 
\end{equation}
where $|\Phi(E_f)\rangle$ is a so-called channel function (with proper antisymmetrization)
given in general by the fragment bound states times their relative free motion and
$V$ is the sum of potentials acting between particles belonging to different fragments.
Thus the transition matrix element $T_{0f,i}$ (additional quantum numbers are dropped)
takes the following form
\begin{eqnarray}
T_{0f,i} &=& \langle \Psi(E_f) | \Theta_i | 0 \rangle \nonumber \\
 &=& \langle \Phi(E_f) | \Theta_i | 0 \rangle + \langle \Phi(E_f) | V (E_f-H+i\eta)^{-1} 
    | 0 \rangle \,.
\end{eqnarray}
The first term of the right-hand-side is the so-called Born term ($T^{\rm Born}_{0f,i}$), 
which can be evaluated without greater problems. The second term ($T^{\rm FSI}_{0f,i}$) 
depends on the final state interaction and its evaluation is much more difficult. However, 
using the LIT approach, 
one can proceed as follows. One rewrites $T^{\rm FSI}_{0f,i}$ in a spectral representation,
\begin{equation}
T^{\rm FSI}_{0f,i} 
 = \sum_n (E_f-E_n) F_{0f,i}(E_f,E_n) + \int_{E_{th}}^\infty (E_f-E'+i\eta)^{-1} F_{0f,i}(E_f,E') dE' 
\end{equation}    
with
\begin{equation}
 F_{0f,i}(E_f,E') = \sum\!\!\!\!~\!\!\!\!\!\!\!\int_{\,\,\Psi_\gamma}
   \langle \Phi(E_f) | V | \Psi_\gamma \rangle 
   \langle \Psi_\gamma | \Theta_i | 0 \rangle \delta(E_f-E') \,.
\end{equation}
The function $F_{0f,i}$ has a similar form as an inclusive response function $R_i$,
therefore one can apply an analogous LIT method as in the inclusive case, however left-
and right-hand sides are not identical, hence two LIT equations are obtained:
\begin{equation} \label{LITeq2}
(H-\sigma_R - i \sigma_I)| \tilde \Psi_i \rangle = \Theta_i |0\rangle \,, \qquad 
(H-\sigma_R - i \sigma_I)| \tilde \Psi_V \rangle = V |\Phi(E_f)\rangle \,.
\end{equation}
The first one is essentially the same as (\ref{LITeq}). The second equation has a different
right-hand side, but with the important feature to vanish asymptotically for a nuclear 
potential $V$. Therefore the equation can again be solved with a bound-state technique. 
In case of an additional Coulomb interaction, one may use Coulomb wave functions instead
of the free motion $|\Phi_f\rangle$ if only two of the fragments carry charge.

Having calculated $\tilde\Psi_i$ and $\tilde\Psi_V$ one evaluates the overlap
$\langle \tilde\Psi_V| \tilde\Psi_i \rangle$,  which is identical to the LIT of the
function $F_{0f,i}$, and hence $F_{0f,i}$ is obtained by the inversion of the LIT. The FSI part
of the T-matrix element is then given by
\begin{equation}
T^{\rm FSI}_{0f,i}(E_f) = -i \pi F_{0f,i}(E_f,E_f) + 
                  \mathcal{P} \int_{E_{th}^-}^\infty (E_f-E')^{-1} F_{0f,i}(E_f,E') dE' 
\end{equation}
and the sum of $T^{\rm FSI}_{0f,i}$ and the simpler Born term $T^{\rm Born}_{0f,i}$
leads to the total result for the transition matrix element.

As shown in \cite{ELOB07} the LIT formalism for exclusive reactions can be reformulated
such that only a solution for $\tilde\Psi_i$ is requested, while $\tilde\Psi_V$ is not
needed. Both possibilities have been used in \cite{Lapiana}, where the $d(e,e'p)n$ reaction
has been calculated as a test case for the exclusive LIT formalism.

\section{Application of the LIT method}

As we have shown in the previous sections the main point of the LIT approach consists
in the fact that a scattering state problem is reduced to a bound-state-like
problem. In other words the calculation of continuum wave functions is not 
required, instead one has to solve equations which can be solved with bound-state techniques.
For A$>$2 the calculation of continuum wave functions is difficult or today even 
impossible, thus, with the LIT method, one can extend the range of calculations to considerably 
larger A. In fact one may conclude the following: if one is able to carry out a bound-state 
calculation for a given particle system then the LIT approach opens up the possibility to 
perform calculations for continuum reactions with this particle system.
In principle one is not restricted to use a specific bound-state technique, but in most
LIT calculations an expansion of ground state $|0\rangle$ and LIT function 
$|\tilde\Psi\rangle$ in hyperspherical harmonics (HH) is employed. Information concerning
such expansions is given in \cite{ELOB07}, here we only want to mention that the 
realistic LIT applications for A=3   
have been performed with the CHH technique \cite{CHH}, whereas the realistic (semirealistic)
applications for A=4 (A$>$4) have been carried out with the EIHH approach \cite{EIHH}.

For the solution of the LIT equation (\ref{LITeq}) the Lanczos method is used in
most cases \cite{Lanczos}. In this context it should be pointed out that the LIT method is 
different from an approach where a so-called Lanczos response $R_{\rm Lanc}$ is introduced, 
which is essentially a LIT with small $\sigma_I$, which, however, is directly interpreted -- 
without any inversion -- as a response function (for more details see \cite{ELOB07}).

\subsection{Simple example: deuteron photodisintegration}

In order to illustrate how the method works we first apply the LIT approach
to a very simple physical problem, namely to the total deuteron photoabsorption cross section
in unretarded dipole approximation. In this case the cross section is given by
\begin{equation}
\sigma_\gamma^d(\omega)= 4 \pi^2 \alpha \omega R_\gamma^d(\omega)\,,
\end{equation}
where $\alpha$ is the fine structure constant, $\omega$ is the energy
of the photon absorbed by the deuteron, and $R_\gamma^d(\omega)$ denotes
the response function defined as
\begin{equation}
   R_\gamma^d(\omega)=\sum\!\!\!\!~\!\!\!\!\!\!\!\!\int _f\,\,
   |\langle f|\Theta| 0 \rangle |^2 \delta(\omega -E_{np}-E_d) \, .
\end{equation}

\noindent Here $E_d$ and $|0\rangle$ are the deuteron bound state energy and wave function, while $E_{np}$
and $|f\rangle$ denote relative kinetic energy and wave function of the outgoing $np$ pair 
for a given two-nucleon Hamiltonian $H$:
\begin{equation}
(H + E_d) |0\rangle =0 \,, \qquad   (H - E_{np}) |f\rangle = 0 \,.
\end{equation}
\noindent The transition operator $\Theta$ is defined by
\begin{equation}\label{operator}
\Theta=\sum_{i=1}^2 z_i \tau_i^3\,,
\end{equation}
where $z_i$ and  $\tau_i^3$ are the third components of position and isospin
coordinates of the i-th nucleon. The LIT of $R_\gamma^d$ is given by
\begin{equation}
 {\rm L}_\gamma^d (\sigma_R,\sigma_I)=\sum_{k=1}^3 \langle\tilde\Psi_k|\tilde\Psi_k\rangle=
  \int R_\gamma^d(\omega)\,  {\mathcal L}(\omega,\sigma_R,\sigma_I)\, d\omega\,,
\end{equation}
where $k=1,2,3$ correspond to different partial waves of the final state, namely
$^3P_0$, $^3P_1$, and $^3P_2- ^3F_2$.   

\begin{figure}[htb]
\begin{center}
\resizebox*{9cm}{5.5cm}{\includegraphics*[angle=0]{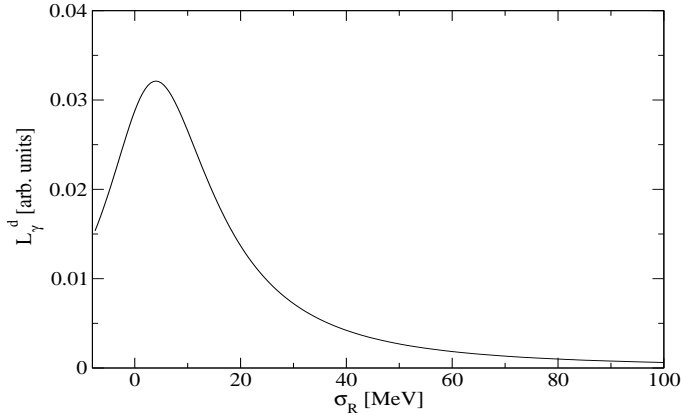}}
\end{center}
\vspace*{8pt}
\caption{LIT L$_\gamma^d$ with $\sigma_I$=10 MeV.}
\end{figure}

\begin{figure}[htb]
\begin{center}
\resizebox*{9cm}{7.5cm}{\includegraphics*[angle=0]{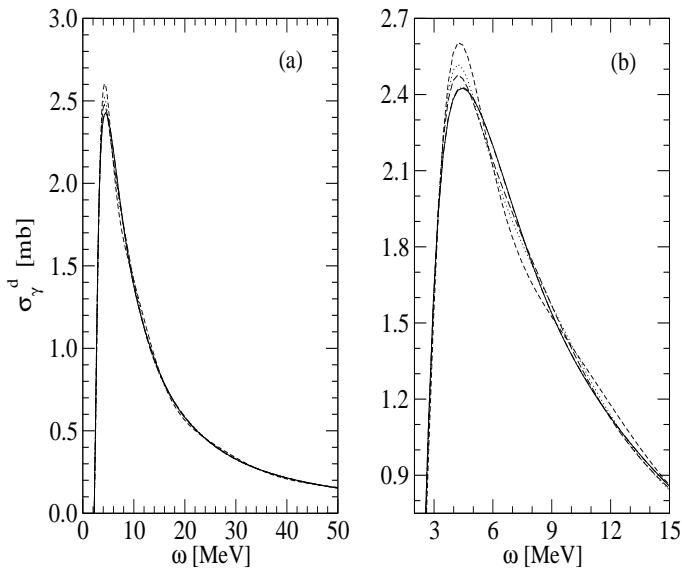}}
\end{center}
\vspace*{8pt}
\caption{$\sigma_\gamma^d$ from inversion of L$_\gamma^d$ of Fig.~1, up to
50 MeV (a) and in peak region (b), with various $M_{\rm max}$ values:
10 (short dashed), 15 (dotted), 20 (long dashed), 25 (solid), 26 (dash-dotted).}
\end{figure}

\begin{figure}[htb]
\begin{center}
\resizebox*{9cm}{7.5cm}{\includegraphics*[angle=0]{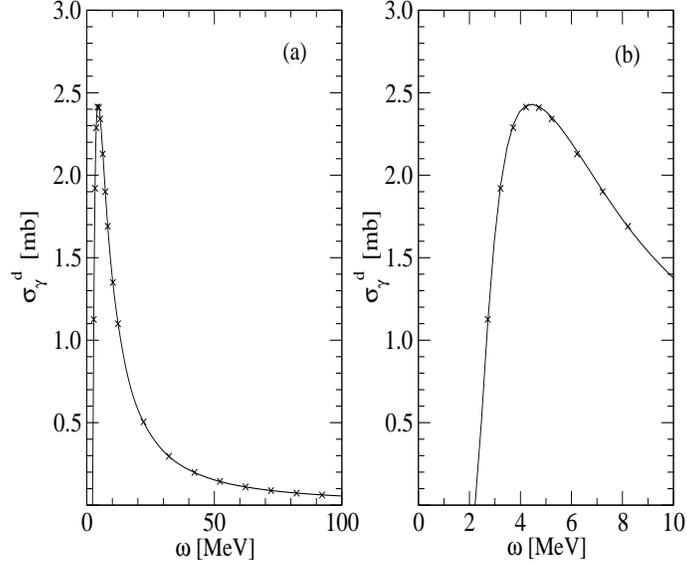}}
\end{center}
\vspace*{8pt}
\caption{Total deuteron photoabsorption cross section up to
50 MeV (a) and in peak region (b): LIT result (solid) and from calculation with
explicit $np$ continuum wave functions (crosses).}
\end{figure}

First we consider deuteron photodisintegration with a realistic NN interaction.
In the already mentioned review article of the LIT method \cite{ELOB07}, such a case
has been investigated using the AV14 NN potential \cite{AV14}. The L$_\gamma^d$ result is
shown in Fig.~1, while the corresponding inversion results are illustrated in Fig.~2.
For the inversion one observes a nice stability range of the results for all the shown 
$M_{\rm max}$ values in the whole considered energy range, except for the peak region, 
where the inversion becomes stable only for higher $M_{\rm max}$. One notes that the
$M_{\rm max}$ values 25 and 26 lead essentially to identical results.

In Fig.~3 the final inversion result ($M_{\rm max}$=25) is compared with the corresponding 
cross section of a conventional calculation with explicit $np$ continuum wave 
functions \cite{ELOB07}. One finds an excellent agreement between the two calculations 
showing that one can reach high-precision results with the LIT method.

\begin{figure}[htb]
\begin{center}
\resizebox*{10cm}{4cm}{\includegraphics*[angle=0]{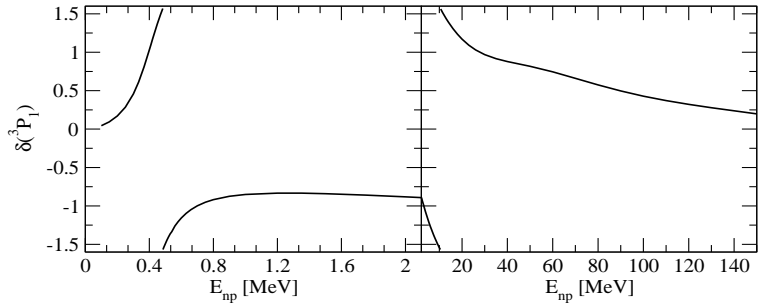}}
\end{center}
\vspace*{8pt}
\caption{Phase shift $^3P_1$ of fictitious $np$ system at low (left) and higher (right)
energies.}
\end{figure}

\begin{figure}[htb]
\begin{center}
\resizebox*{10cm}{4cm}{\includegraphics*[angle=0]{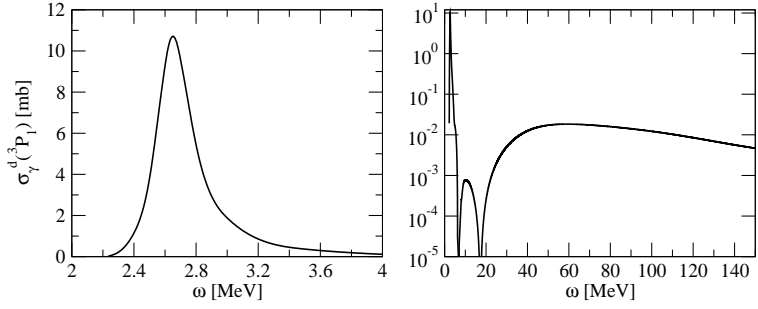}}
\end{center}
\vspace*{8pt}
\caption{As Fig.~4 but for photodisintegration cross section.}
\end{figure}

Further LIT calculations for the deuteron total photoabsorption cross sections have already been
discussed in \cite{JISP,WL08}. For the aim of the present discussion, {\it i.e.} the way of 
working of the LIT method, \cite{WL08} is particularly interesting. It is a case study for a 
fictitious $np$ system with a low-lying and narrow resonance in the $^3P_1$ 
nucleon-nucleon partial wave (obtained by an additional attractive term, for details see
\cite{WL08}). The results of a conventional calculation with the fictitious $np$ system 
for the $^3P_1$ phase shifts and the ``deuteron photoabsorption cross section'' to the $^3P_1$ 
final state are shown in Figs.~4 and 5, respectively. The $^3P_1$ phase shifts exhibit two 
resonances, at $E_{np}=0.48$ MeV and at about $E_{np}=10.5$ MeV. The low-energy resonance leads 
to the dominant structure of the photoabsorption cross section, a pronounced peak at a 
photon energy of 2.65 MeV with a width $\Gamma$ of 270 keV, while the second resonance only 
shows up as a rather tiny peak, which is more than four orders of magnitude smaller than the 
first peak.

\begin{figure}[htb]
\begin{center}
\resizebox*{6cm}{6.5cm}{\includegraphics*[angle=0]{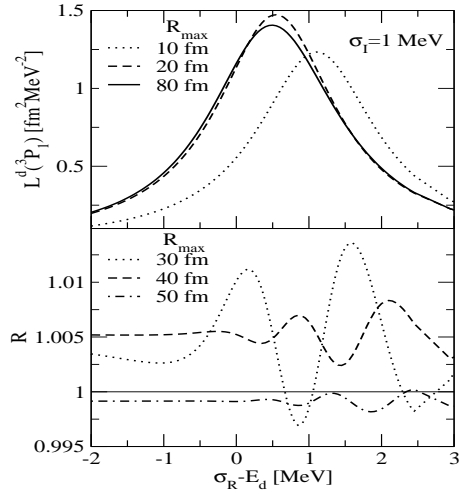}}
\end{center}
\vspace*{8pt}
\caption{LIT L$^d_\gamma(^3P_1)$ of fictitious $np$ system in
first resonance region with $\sigma_I=1$ MeV and various values of $R_{\rm max}$ (top) and ratio
R=L$^d_\gamma(R_{\rm max})/L^d_\gamma(R_{\rm max}=80$ fm) (bottom).}
\end{figure}

For the LIT calculation of the photoabsorption cross section of the fictitious $np$ system
the inversion basis set $\chi_m$ (\ref{bsetn}) is modified to account for the resonant 
structure, to this end the functions $\chi_m$ are relabeled: 
$\chi_m \rightarrow \chi_{m+1}$. In addition a new $\chi_1$ is defined,
\begin{equation}
\label{bset1}
\chi_1(E_{np},\alpha_i) = {\frac {1} {(E_{np} -E_{\rm res})^2 + ({\frac {\Gamma}{2}})^2}}
\left({\frac {1}{1+\exp(-1)}}-{\frac {1}{1+\exp((E_{np}-\alpha_3)/\alpha_3)}}\right)\,,
\end{equation}
where $E_{\rm res} \equiv \alpha_4$, $\Gamma \equiv \alpha_5$ and $\alpha_3$ are additional 
nonlinear parameters. 

\begin{figure}[htb]
\begin{center}
\resizebox*{11.5cm}{7.5cm}{\includegraphics*[angle=0]{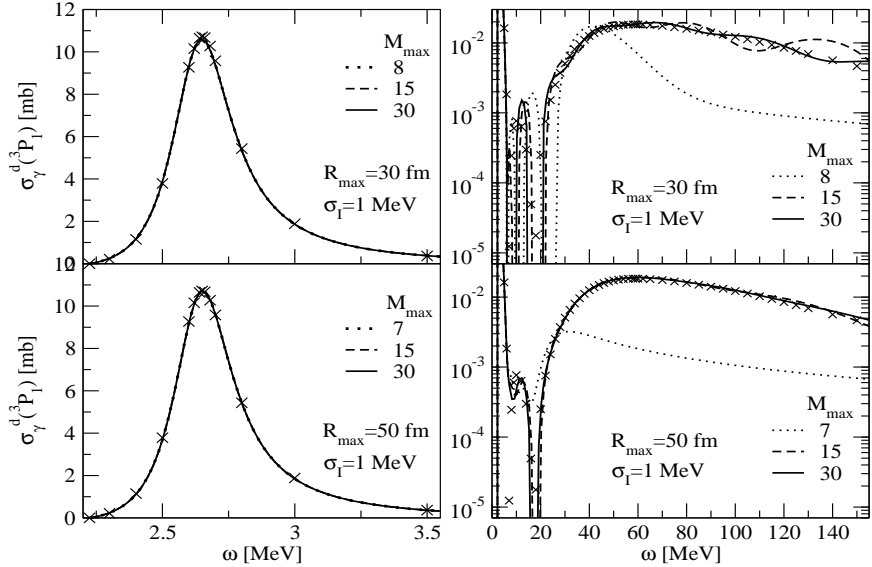}}
\end{center}
\vspace*{8pt}
\caption{$\sigma^d_{\gamma}(^3P_1)$ of fictitious $np$ system from inversion of
L$^d_\gamma(^3P_1, \sigma_I=1$ MeV, $R_{\rm max})$ with various $M_{\rm max}$ values and
$R_{\rm max}=30$ fm (top) and 50 fm (bottom) in first resonance region (left) and beyond (right);
also shown results of a conventional calculation with explicit $np$ continuum wave functions
(crosses).}
\end{figure}

The reason for such a LIT case study with a resonance lies in the results of a previous LIT 
calculation for the (e,e') longitudinal and transverse form factors of $^4$He \cite{Sonia07}, 
where a resonance in the Coulomb monopole transition was obtained, but its width 
could not be determined. In the case study it is shown that for a proper 
resolution of a resonant structure it is very important to take into account the LIT function 
$\tilde\Psi$ up to rather large distances \cite{WL08}. This has been checked (i) by solving 
(\ref{LITeq}) imposing at a two-nucleon distance $r=R_{\rm max}$ an asymptotic boundary condition
which leads to a strong fall-off of $\tilde\Psi(^3P_1)$ and (ii) by calculating the norm
$\langle\tilde\Psi(^3P_1)|\tilde\Psi(^3P_1)\rangle$ only in
the range from $r=0$ to $r=R_{\rm max}$. In Fig.~6 we show the results for such
a calculation choosing $\sigma_I=1$ MeV. One notes that for a rather precise result,
with errors below about 1\%, one has to take $R_{\rm max}\ge 30$ fm. A further increase
of $R_{\rm max}$ to 50 fm leads to a reduction of the relative error by about a factor ten.
In Fig.~7 the inversions results with $R_{\rm max}=30$ and 50 fm are depicted in comparison to
the result of the direct calculation. One observes that for both $R_{\rm max}$ values
the resonance cross section is described with high accuracy. Differences between the two
cases become evident in the region of the second maximum and at higher energy. In fact
with $R_{\rm max}=30$ fm one finds only a reasonably good description, while
a considerable improvement is obtained with $R_{\rm max}=50$ fm.

\begin{figure}[htb]
\begin{center}
\resizebox*{6cm}{6.5cm}{\includegraphics*[angle=0]{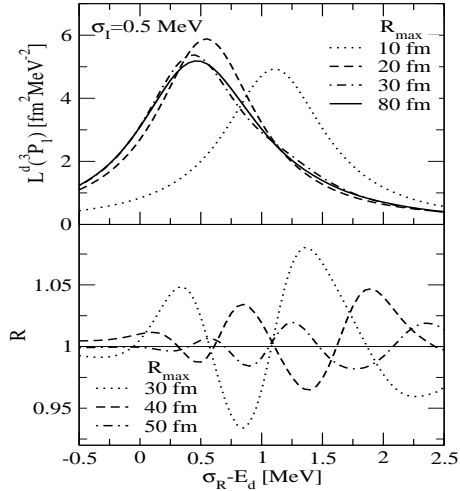}}
\end{center}
\caption{As Fig.~6 but with $\sigma_I=0.5$ MeV
}
\end{figure}

As next point we consider the reduction of $\sigma_I$ from 1 MeV to 0.5 MeV. As shown in
Fig.~8 one finds an enhancement of the relative error of $L_\gamma^d$ by at least a factor 
of five in comparison to the corresponding $R_{\rm max}$ values of Fig.~6. The enhancement 
is easily understood by investigating the asymptotic solution of (\ref{LITeq}); in the 
here considered deuteron case it is described by the exponential fall-off 
$\exp(-r(M\sigma_I)^{\frac{1}{2}}/\hbar)$, where $M$ is the nucleon mass. It is evident that a 
smaller $\sigma_I$ leads to a longer range LIT function $\tilde\Psi$. As discussed in 
\cite{WL08}, for $\sigma_I=0.1$ MeV even $R_{\rm max}=$300 fm is not completely sufficient, 
since only the resonance itself is resolved, but not the cross section at higher energies. 

The discussion above seems to infer that it is better to choose a rather large value for 
$\sigma_I$. On the other hand it should also be clear that the width of the resonance $\Gamma$ 
and the value of $\sigma_I$ are correlated. If $\sigma_I$ is too large the resonance cannot 
be resolved. In the case study it has been found that $\sigma_I=2$ MeV, about seven times
larger than the width $\Gamma$, is still sufficient for a resolution of the resonance. However, 
in a general case the resonance width is not known beforehand and the question arises 
what should be the proper value for $\sigma_I$ in such a case. As pointed out in \cite{WL08}
one has to proceed as follows. One performs a LIT calculation with a given $\sigma_I$ and 
compares the result to a LIT with a $\delta$-peak in the response function or cross section.
For example, in the here discussed deuteron case one sets
\begin{equation}
\nonumber
R_\gamma^{\rm peak}(E_{np})= R^{\rm peak} \delta(E_{np}-E^{\rm peak}) \,.
\end{equation}
The resulting $\delta$-LIT is then given by the Lorentzian function
\begin{equation}
\nonumber
{\rm L}^d_\delta(E^{\rm peak},\sigma_R,\sigma_I) = R^{\rm peak}
{\mathcal L}(E^{\rm peak},\sigma_R,\sigma_I) \,,
\end{equation}
where $R^{\rm peak}$ is chosen such that the peak heights of ${\rm L}^d_\delta$ and
${\rm L}^d_\gamma$ are equal.

\begin{figure}[htb]
\begin{center}
\resizebox*{9cm}{7cm}{\includegraphics*[angle=0]{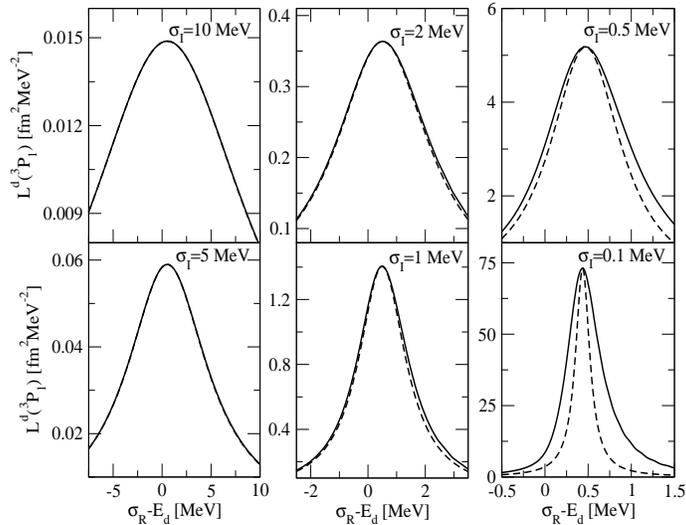}}
\end{center}
\vspace*{8pt}
\caption{LIT L$^{\rm d}(^3P_1)$ (full curves) and L$^{\rm d}_\delta$ (dashed curves) of 
fictitious $np$ system in first resonance region  with various $\sigma_I$ values
and $R_{\rm max}=80$ fm, except for $\sigma_I=0.1$ MeV where $R_{\rm max}$=300 fm.
}
\end{figure}

In order to obtain a reliable inversion the actual LIT ${\rm L}^d_\gamma$ should have a 
larger width than the $\delta$-LIT ${\rm L}^d_\delta$. If, on the contrary, they lead to 
essentially identical results, one has to reduce $\sigma_I$ up to the point that the actual 
LIT is sufficiently different from the corresponding $\delta$-LIT. In Fig.~9 we show such 
results for the deuteron case study. For $\sigma_I$=5 and 10 MeV there are practically no 
differences between LIT and $\delta$-LIT, while for $\sigma_I \le 2$ MeV 
differences become visible. As a matter of fact $\sigma_I$=2 MeV is sufficient for 
a reliable inversion and thus one may conclude that in a general case one has to use a
$\sigma_I$ such that  differences between LIT and $\delta$-LIT have at least the same size
as in  the $\sigma_I$=2 MeV case of Fig.~9.

In Fig.~10 we show the final inversion results from \cite{WL08} with $R_{\rm  max}=$80 fm. 
One sees that the E1 resonance is precisely described with the LIT method for $\sigma_I=0.5$ 
and 2 MeV, while the peak is somewhat underestimated with $\sigma_I=5$ MeV. In the resonance 
region, with the two lower $\sigma_I$ values, essentially the same results are obtained 
as for the cases of Fig.~7 with $\sigma_I$=1 MeV and $R_{\rm  max}=$30 and 50 fm. From the 
comparison one further notes that the case $\sigma_I$=2 MeV and 
$R_{\rm  max}=80$ fm leads to even more precise results in the second resonance region, and 
beyond, than shown in Fig.~7 for $R_{\rm  max}=$50 fm. 

\begin{figure}[htb]
\begin{center}
\resizebox*{11cm}{5.5cm}{\includegraphics*[angle=0]{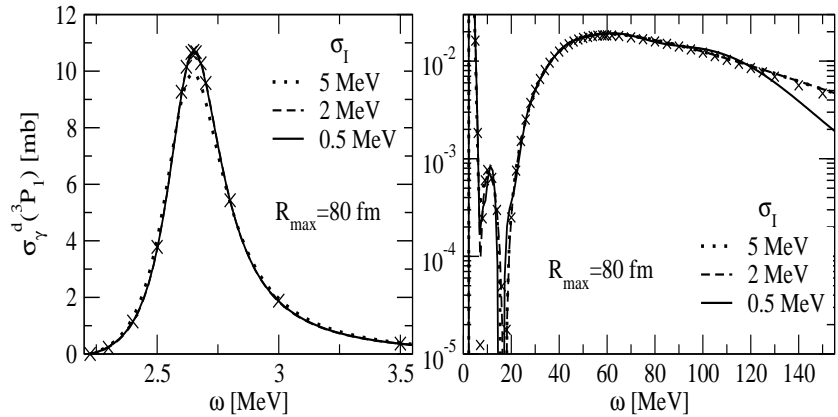}}
\end{center}
\vspace*{8pt}
\caption{As Fig.~7 but with $R_{\rm max}=80$ fm and $\sigma_I=5$, 2, and 0.5 MeV.
}
\end{figure}

\subsection{Reactions with $A \ge 3$}

Now we turn to realistic applications of the LIT method and consider first the $^4$He total
photoabsorption cross section. In fact this is one of the very first LIT 
applications \cite{4-body}, however, initially only performed with semirealistic NN forces. 
In \cite{4-body} it has been found that the calculated cross section shows a considerably 
more pronounced giant dipole resonance than the most recent experimental data of that time. 
Such a difference between experiment and an {\it ab initio} calculation has led to a renewed 
experimental interest in the $^4$He photoabsorption cross section. In fact in the meantime 
three additional experiments have been carried out \cite{Nilsson,Shima,Nakayama}. Finally, 
in \cite{Doron06} the first calculation with realistic NN and 3N forces has also been 
published. In Fig.~11 we show the theoretical results of \cite{Doron06} together with the 
experimental data. One notes that the 3N force leads to a considerable reduction of the peak 
cross section. One further sees that there is a very good agreement between theory and the 
data of \cite{Nakayama} and also quite a good agreement with the data of \cite{Nilsson}, 
while the cross section of \cite{Shima} shows a completely different behavior. Also shown 
in the figure are data from experiments performed about 20 years ago. 

\begin{figure}[htb]
\begin{center}
\resizebox*{8cm}{6cm}{\includegraphics*[angle=0]{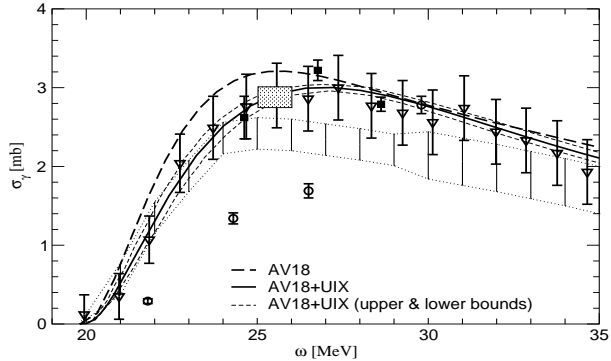}}
\end{center}
\vspace*{8pt}
\caption{Total $^4$He photoabsorption cross section: LIT calculation with AV18 NN potential
only \cite{AV18},
and with additional UIX 3N force \cite{UIX};
experimental data: squares \cite{Nilsson},
circles \cite{Shima},
triangles \cite{Nakayama},
area between dotted lines \cite{Berman,Feldman},
and dotted box \cite{Wells}.
}
\end{figure}

It should be mentioned that today two
other LIT calculations for the $^4$He total photoabsorption cross section are 
available \cite{Sonia_UCOM,Sofia_NCSM}, where different realistic nuclear forces have been used. 
Essentially, they confirm the results of \cite{Doron06}.

In Fig.~12 LIT results for the $^6$Li and $^6$He total photoabsorption cross sections
calculated with semirealistic NN forces are shown \cite{6-body}. For $^6$Li one finds a single
and rather broad cross section peak. On the contrary for $^6$He a very interesting cross section
with a double peak becomes apparent in the calculation. This microscopic result can be
interpreted as follows. In a cluster picture, with an $\alpha$ core and a di-neutron,
the low-energy peak is due to the relative motion of di-neutron and $\alpha$ core. The second 
peak, however, cannot be obtained in a cluster model, but is explained by the classical E1 
giant resonance picture with a collective response of all nucleons (relative motion of protons 
and neutrons). For
$^6$Li, in a cluster model described by an $\alpha$ core plus a deuteron, a similar low-energy
peak is missing, because the deuteron knock-out corresponds to an isoscalar transition,
which cannot be induced by the isovector dipole operator (\ref{operator}). On the other hand
a transition to the antibound $^1S_0(np)$ plus $\alpha$ core is possible. The nucleus
$^6$Li exhibits a considerably larger width of the giant dipole peak than $^6$He. In fact in
the former a break-up into two three-body nuclei is possible ($^3$H$-^3$He), while a similar
break-up of $^6$He is not induced by the dipole operator, since the $^3$H$-^3$H pair has no
dipole moment. The experimental $^6$He and $^6$Li photoabsorption cross sections are not yet
well settled (see \cite{6-body}) and therefore not shown here.

\begin{figure}[htb]
\begin{center}
\resizebox*{7cm}{7cm}{\includegraphics*[angle=0]{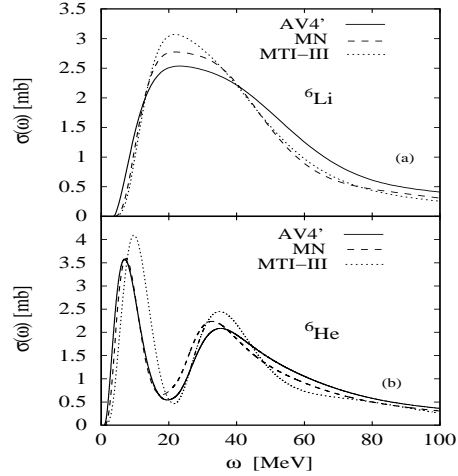}}
\end{center}
\vspace*{8pt}
\caption{$^6$Li and $^6$He total photoabsorption cross sections with various
semirealistic NN potential models: AV4' \cite{AV4'}, 
MN \cite{MN}, 
and MTI-III \cite{MT}.
}
\end{figure}

\begin{figure}[htb]
\begin{center}
\resizebox*{7cm}{4.5cm}{\includegraphics*[angle=0]{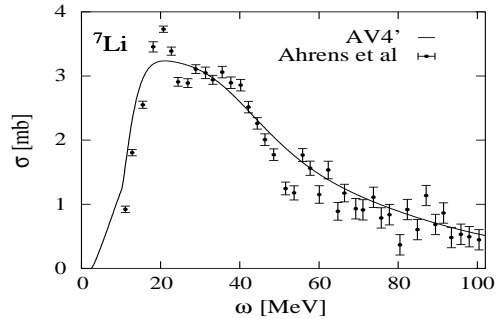}}
\end{center}
\vspace*{8pt}
\caption{$^7$Li total photoabsorption cross section with semirealistic NN potential model
AV4', experimental data from \cite{Ahrens}.
}
\end{figure}

In Fig.~13 we depict the LIT calculation for the $^7$Li total photoabsorption cross 
section \cite{7-body}, in comparison to experimental data \cite{Ahrens}. It is worthwhile to 
mention that the experimental cross section
has not been determined by summing up the various channel cross sections, but in a single
experiment by the ``diminution of photon flux'' method. Like $^6$Li also $^7$Li has a giant 
dipole resonance peak with a rather large width. The comparison of experimental and
theoretical results shows quite a good agreement, though only a semirealistic NN force 
has been used in the LIT calculation. Of course, it would be very interesting to have even 
more precise data and also a calculation with realistic nuclear forces.

\begin{figure}[htb]
\begin{center}
\resizebox*{12cm}{9cm}{\includegraphics*[angle=0]{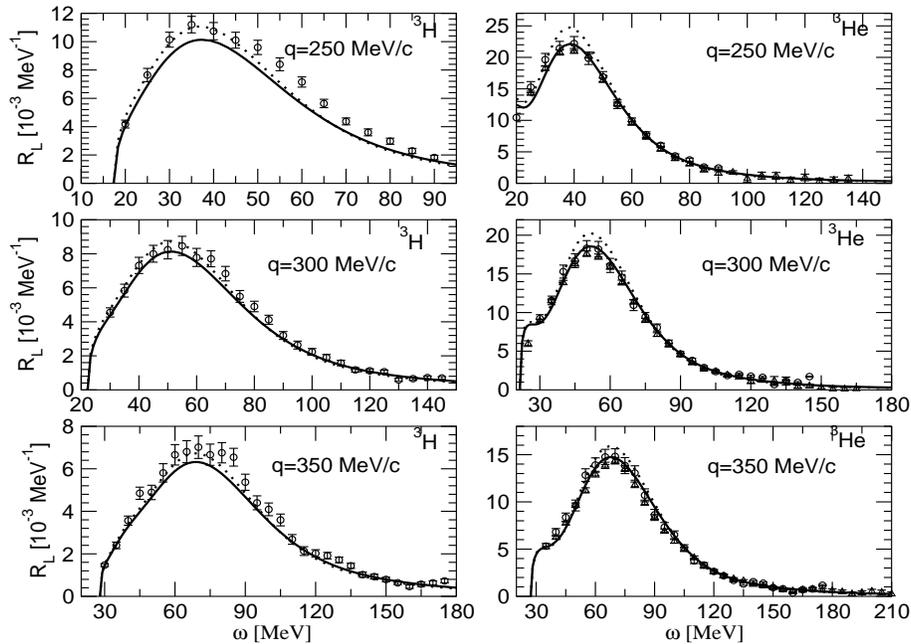}}
\end{center}
\vspace*{8pt}
\caption{$R_L(\omega,q)$ of $^3$H (left) and $^3$He (right)
at various $q$: AV18 NN potential (dotted), AV18 NN + UIX 3N potential
(solid); experimental data: triangles \cite{Saclay}, 
and circles \cite{Bates}. 
}
\end{figure}

\begin{figure}[htb]
\begin{center}
\resizebox*{8.5cm}{8.5cm}{\includegraphics*[angle=-90]{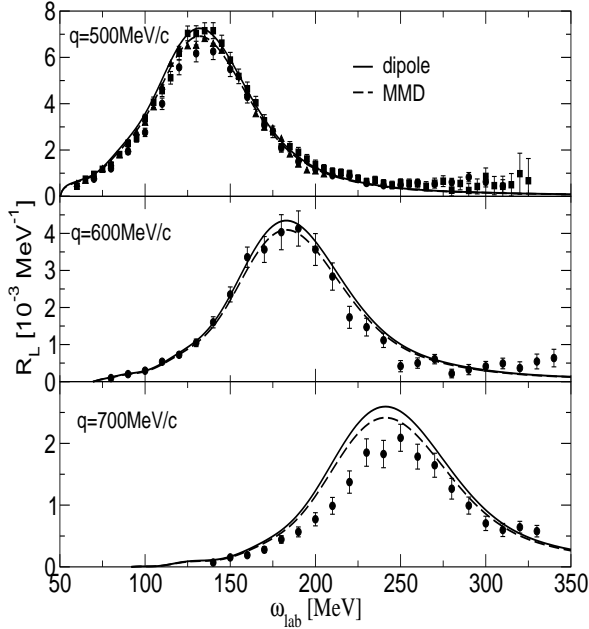}}
\end{center}
\vspace*{8pt}
\caption{$R_L(\omega,q)$ of $^3$He at various $q$, calculation with AV18 NN + UIX 3N 
potentials and two nucleon form factors fits: dipole (solid) and MMD from \cite{MMD} 
(dashed); experimental data: squares \cite{Saclay}, 
triangles \cite{Bates},
and circles \cite{world}.
}
\end{figure}

Now we turn to the inclusive electron scattering response functions. For $^3$H and $^3$He LIT 
calculations for the longitudinal response function $R_L(\omega,q)$ have been carried out
with realistic nuclear forces for momentum transfers below \cite{ee'04}, and above 
$q$=500 MeV/c \cite{ee'05}. Relativistic corrections for the transition operator $O_L$ have 
been taken into account and the frame dependence of the essentially nonrelativistic calculation 
has been studied. In Fig.~14 $R_L$ is shown for various lower $q$ values \cite{ee'04}.
One notes that the 3N force reduces the quasielastic peak height somewhat. The 3N force effect, 
however, does not lead to a consistent picture in comparison with experiment. In fact one 
finds an improvement for $^3$He and a deterioration for $^3$H. 

In \cite{ee'05} it has been found that relativistic effects due to 
the kinetic energy can be largely reduced if $R_L$ is first calculated in a specific
reference frame, where the target nucleus moves with --A${\bf q}/2$, and then transformed to the 
lab system. Different from a direct calculation in the lab system, one finds a 
correct result for the experimentally established quasielastic peak position \cite{ee'05}, 
as shown in Fig.~15. At $q$=500 and 600 MeV/c also for the peak height a good agreement with 
experimental data is obtained, whereas the theoretical $R_L$ overestimates the experimental 
one at $q$=700 MeV/c (at even higher $q$ experimental data are not available). As Fig.~15 shows 
also the choice of the nucleon form factor fit has a non-negligible impact on the result, 
but cannot explain the discrepancy with the data at $q=700$ MeV/c.

\begin{figure}[htb]
\begin{center}
\resizebox*{7cm}{12cm}{\includegraphics*[angle=0]{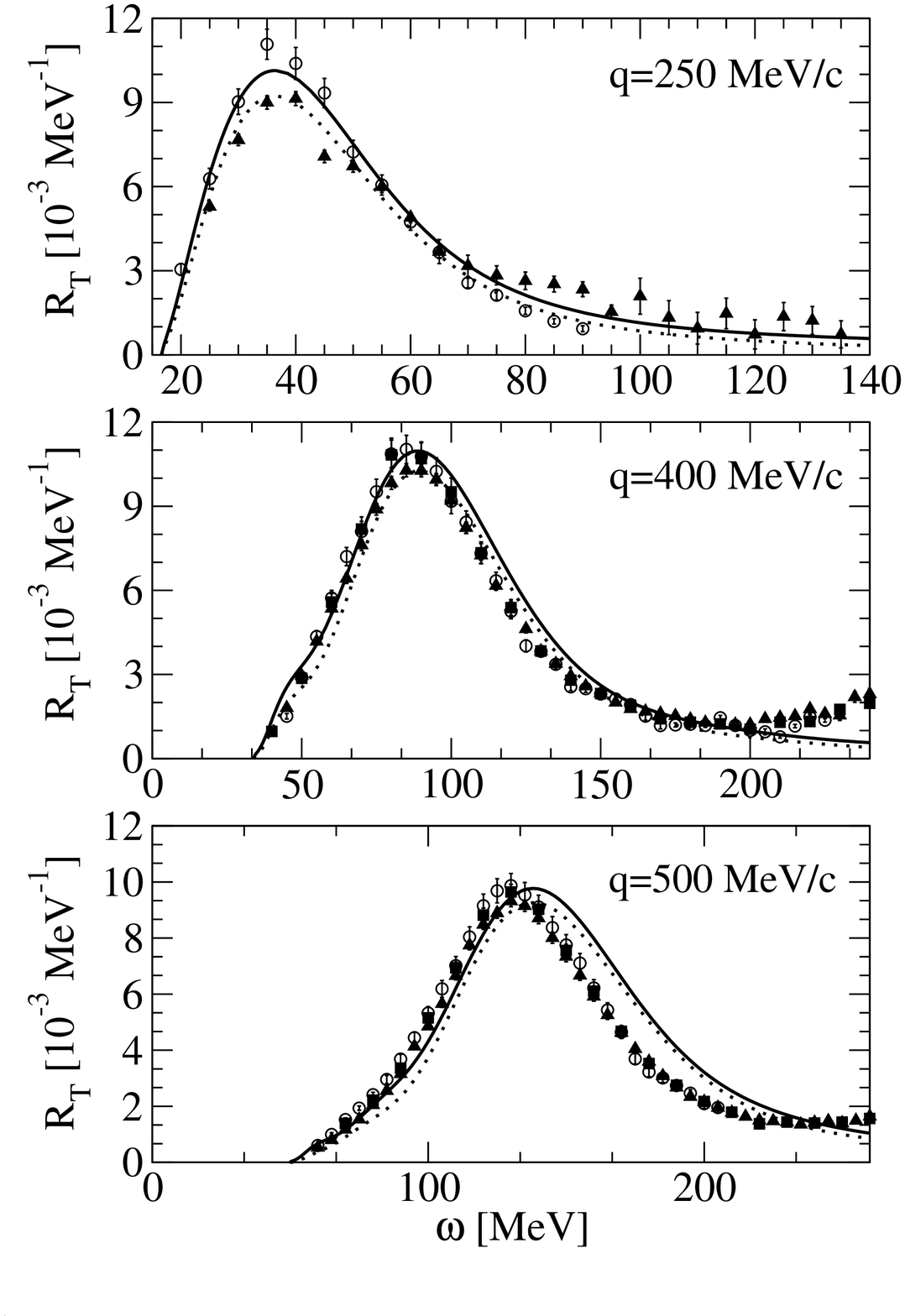}}
\resizebox*{7cm}{12cm}{\includegraphics*[angle=0]{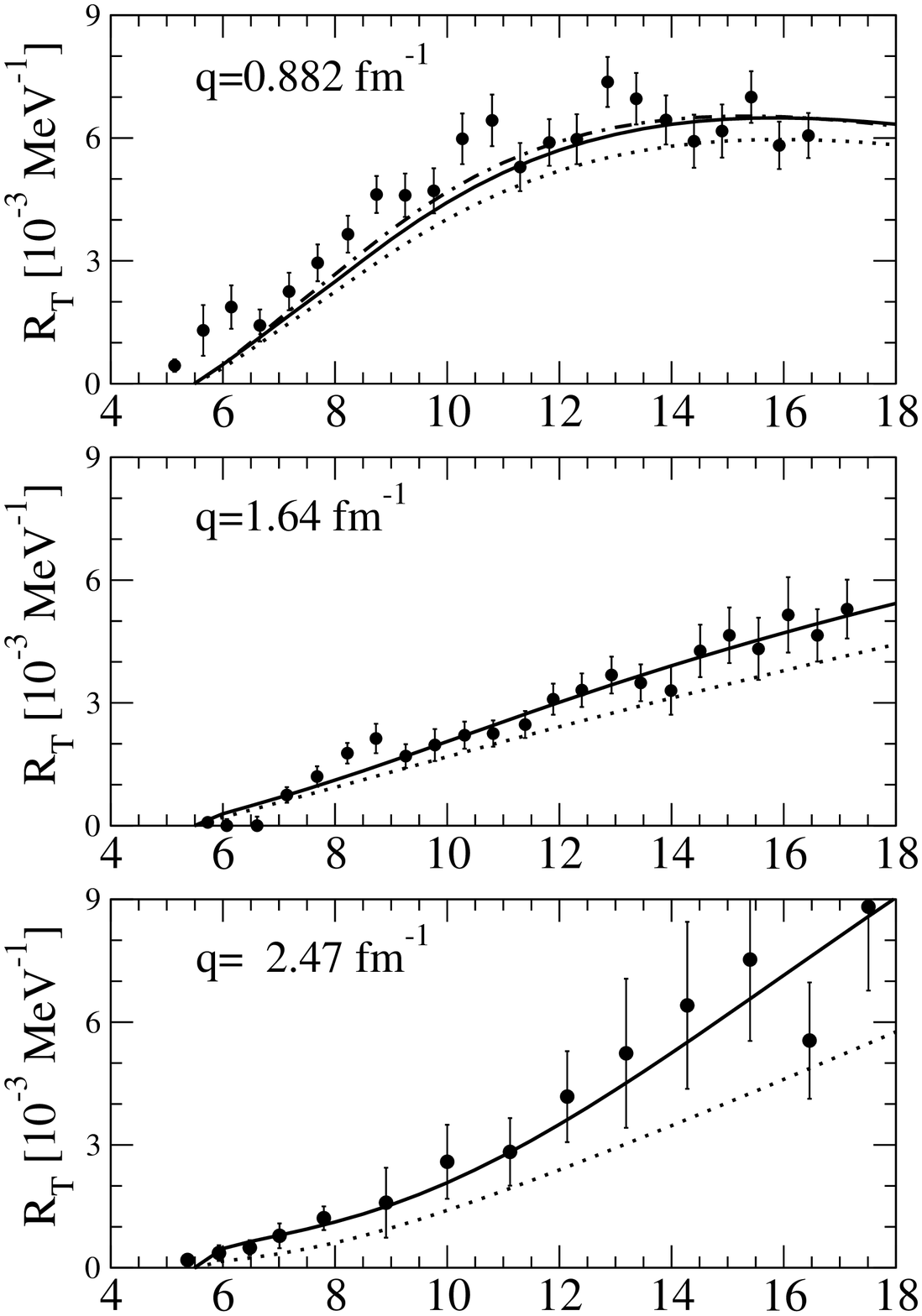}}
\end{center}
\vspace*{8pt}
\caption{$R_T(\omega,q)$ in quasielastic (left) and threshold region (right), 
theoretical results with BonnRA NN potential \cite{Bonn}, 
and TM' 3N forces \cite{TM}, 
and following current operators: one-body (dotted) and one-body + $\pi$-EC +
$\rho$-EC + additional EC via Siegert operator (solid). Experimental data: left panels,
triangles \cite{Saclay}, 
circles \cite{Bates}, 
and squares \cite{world}; 
and right panels, circles \cite{Retzlaff}.
}
\end{figure}

Recently we also calculated the transverse response function $R_T(\omega,q)$ with 
realistic nuclear forces ($q \le 500$ MeV/c) \cite{ee'08}. Besides the usual one-body 
operators also $\pi$ and $\rho$ exchange currents (EC), consistent with the NN potential model, 
were taken into account. In addition also the effect of the so-called Siegert operator
has been studied. In Fig.~16 the theoretical results are shown together with experimental data. 
In the quasielastic region EC lead to some increase of the peak height 
(see left panels of Fig.~16), but the EC effect is much larger close
to threshold (see right panels of Fig.~16) and is important for a good agreement of 
theory and experiment. For $q$=500 MeV/c one finds different peak positions in theory
and experiment, presumably due to the fact that $R_T$ is calculated directly in the lab frame
(see discussion above). The frame dependence of $R_T$ is presently under investigation.

Presently we are extending our realistic calculation to the $(e,e')$
response functions of $^4$He and results for $R_L$ will be published soon.

Summarizing one may conclude that the LIT method is very powerful and allows
a considerable extension of the range of microscopic {\it ab initio} calculations.


\newpage

\begin{thebibliography}{00}
\bibitem{ELO94} V. D. Efros, W. Leidemann, and G. Orlandini, {\it Phys. Lett.} 
 {\bf B338}, 130 (1994).
\bibitem{Efr85} V. D. Efros, {\it Yad. Fiz.} {\bf 41}, 1498 (1985) 
 [{\it Sov. J. Nucl. Phys.} {\bf 41}, 949 (1985)].
\bibitem{Stieltjes} V. D. Efros, W. Leidemann, and G. Orlandini, {\it Few-Body Syst.} 
 {\bf 14}, 151 (1993).
\bibitem{ELOT00} V. D. Efros, W. Leidemann, G. Orlandini, and E. L. Tomusiak,
 {\it Phys. Lett.} {\bf B484}, 223 (2000).
\bibitem{Doron06} D. Gazit {\it et al.}, {\it Phys. Rev. Lett.} {\bf 96}, 112301 (2006).
\bibitem{GN07} D. Gazit and N. Barnea, {\it Phys. Rev. Lett.} {\bf 98}, 192501 (2007).
\bibitem{4-body} V. D. Efros, W. Leidemann, and G. Orlandini, {\it Phys. Rev. Lett.} {\bf 78}, 
 4015 (1997); N. Barnea, V. D. Efros, W. Leidemann, and G. Orlandini, {\it Phys. Rev.} {\bf C63}, 
 057002 (2001). 
\bibitem{6-body} S. Bacca {\it et al.}, {\it Phys. Rev. Lett.} {\bf 89}, 052502 (2002); S. Bacca,  N. Barnea, W. Leidemann, and G. Orlandini, {\it Phys. Rev.} {\bf C69}, 057001 (2004).
\bibitem{7-body} S. Bacca {\it et al.}, {\it Phys. Lett.} {\bf B603}, 159 (2004).
\bibitem{ee'97} V. D. Efros, W. Leidemann, and G. Orlandini, {\it Phys. Rev. Lett.}
 {\bf 78}, 432 (1997).
\bibitem{ee'05} V. D. Efros, W. Leidemann, G. Orlandini, and E. L. Tomusiak, {\it Phys. Rev.} 
{\bf C72}, 011002(R) (2005).
\bibitem{ee'08} S. Della Monaca {\it et al.}, {\it Phys. Rev.} {\bf C77}, 044007 (2008).  
\bibitem{Sofia1} S. Quaglioni {\it et al.}, {\it Phys. Rev.} {\bf C69}, 044002 (2004).
\bibitem{Sofia2} S. Quaglioni, V. D. Efros, W. Leidemann, and G. Orlandini,
 {\it Phys. Rev.} {\bf C72}, 064002 (2005).
\bibitem{Diego} D. Andreasi {\it et al.}, {\it Eur. Phys. J.} {\bf A27}, 47 (2006).
\bibitem{ELOB07} V. D. Efros, W. Leidemann, G. Orlandini, and N. Barnea,
 {\it J. Phys.} {\bf G34}, R459 (2007).
\bibitem{textbook} J. D. Walecka, {\it Theoretical Nuclear and Subnuclear Physics} 
 (World Scientific, Singapore, 2004).
\bibitem{pol_deuteron} H. Arenh\"ovel, W. Leidemann, and E. L. Tomusiak,
 {\it Eur. Phys. J.} {\bf A23}, 147 (2005).
\bibitem{Diego2} D. Andreasi, W. Leidemann, Ch. Reiss, and M. Schwamb,
  {\it Eur. Phys. J.} {\bf A24}, 361 (2005).
\bibitem{Goldb} M. L. Goldberger and K. W. Watson, {\it Collision Theory} (Wiley, New York, 1964).
\bibitem{Lapiana} A. La Piana and W. Leidemann, {\it Nucl. Phys.} {\bf A677}, 423 (2000).
\bibitem{CHH} Yu. I. Fenin and V. D. Efros, {\it Yad. Fiz.} {\bf 15}, 887 (1972) 
 [{\it Sov. J. Nucl. Phys.} {\bf 15}, 497 (1972)]; W. Leidemann, V. D. Efros, G. Orlandini, 
 and E. L. Tomusiak, {\it Fizika} {\bf B8}, 135 (1999).
\bibitem{EIHH} N. Barnea, W. Leidemann, and G. Orlandini, {\it Phys. Rev.} 
 {\bf C61}, 054001 (2000);
 N. Barnea, W. Leidemann, and G. Orlandini, {\it Nucl. Phys.} {\bf A693}, 565 (2001).
\bibitem{Lanczos} M. Marchisio, N. Barnea, W. Leidemann, and G. Orlandini, 
 {\it Few-Body Syst.} {\bf 33}, 259 (2003).
\bibitem{AV14} R. B. Wiringa, R. A. Smith, and T. L. Ainsworth, {\it Phys. Rev.}
 {\bf C29}, 1207 (1984).
\bibitem{JISP} N. Barnea, W. Leidemann, and G. Orlandini, {\it Phys. Rev.}
 {\bf C74}, 034003 (2006).
\bibitem{WL08} W. Leidemann, {\it Few-Body Syst.} in print, arXiv:0803.1770.
\bibitem{Sonia07} S. Bacca {\it et al.} {\it Phys. Rev.} {\bf C76}, 014003 (2007).
\bibitem{Nilsson} B. Nilsson {\it et al.}, {\it Phys. Lett.} {\bf B626}, 65 (2005).
\bibitem{Shima} T. Shima {\it et al.}, {\it Phys. Rev.} {\bf C72}, 044004 (2005).
\bibitem{Nakayama} S. Nakayama {\it et al.}, {\it Phys. Rev.} {\bf C76}, 021305(R) (2007).
\bibitem{AV18} R. B. Wiringa, V. G. J. Stoks, and R. Schiavilla, {\it Phys. Rev.} 
 {\bf C51}, 38 (1995).
\bibitem{UIX} B. S. Pudliner {\it et al.}, {\it Phys. Rev.} {\bf C56}, 1720 (1997).
\bibitem{Berman} B. L. Berman, D. D. Faul, P. Meyer, and D. L. Olson,
 {\it Phys. Rev.} {\bf C22}, 2273 (1980).
\bibitem{Feldman} G. Feldman {\it et al.}, {\it Phys. Rev.} {\bf C42}, R1167 (1990).
\bibitem{Wells} D. P. Wells {\it et al.}, {\it Phys. Rev.} {\bf C46}, 449 (1992). 
\bibitem{Sonia_UCOM} S. Bacca, {\it Phys. Rev.} {\bf C75}, 044001 (2007).
\bibitem{Sofia_NCSM} S. Quaglioni and P. Navr\`atil, {\it Phys.Lett.} {\bf B652}, 370 (2007).
\bibitem{AV4'} R. B. Wiringa and S. C. Pieper, {\it Phys. Rev. Lett.} 89, 182501 (2002).
\bibitem{MN} D. R. Thomson, M. LeMere, and Y. C. Tang, {\it Nucl. Phys.} {\bf A286},
 53 (1977); I. Reichstein and Y. C. Tang, {\em ibid.} {\bf A158}, 529 (1970).
\bibitem{MT} R. A. Malfliet and J. Tjon, {\it Nucl. Phys.} {\bf A127}, 161 (1969).
\bibitem{Ahrens} J. Ahrens {\it et al.}, {\it Nucl. Phys.} {\bf A251}, 479 (1975).
\bibitem{ee'04} V. D. Efros, W. Leidemann, G. Orlandini, and E. L. Tomusiak, 
 {\it Phys. Rev.} {\bf C69}, 044001 (2004).
\bibitem{Saclay} C. Marchand {\it et al.}, {\it Phys. Lett.} {\bf B153}, 29 (1985).
\bibitem{Bates} K. Dow {\it et al.}, {\it Phys. Rev. Lett.} {\bf 61}, 1706 (1988).
\bibitem{world} J. Carlson, J. Jourdan, R. Schiavilla, and I. Sick,  {\it Phys. Rev.} 
 {\bf C65}, 024002 (2002).
\bibitem{MMD} P. Mergell, U.-G. Meissner, and D. Drechsel, {\it Nucl. Phys.}
 {\bf A596}, 367 (1996).
\bibitem{Bonn} R. Machleidt, {\it Adv. Nucl. Phys.} {\bf 19}, 189 (1989).
\bibitem{TM} S. A. Coon {\it et al.}, {\it Nucl. Phys.} {\bf A317}, 242 (1979).
\bibitem{Retzlaff} G. A. Retzlaff {\it et al.}, {\it Phys. Rev.} {\bf C49}, 1263 (1994).
    
\end{thebibliography}
\end{document}